\documentclass[prb,twocolumn,showpacs,floatfix]{revtex4} 
\usepackage{amsmath,amssymb,isolatin1,graphicx,times}

\newcommand{\be}{\begin{equation}}
\newcommand{\ee}{\end{equation}}
\newcommand{\bea}{\begin{eqnarray}}
\newcommand{\eea}{\end{eqnarray}}
\newcommand{\bw}{\begin{widetext}}
\newcommand{\ew}{\end{widetext}}

\newcommand{\kommentar}[1]{}

\begin{document}
 
\title{Coherent exciton transport in dendrimers and continuous-time
quantum walks}
\author{Oliver M{\"u}lken}
\author{Veronika Bierbaum}
\author{Alexander Blumen}
\affiliation{
Theoretische Polymerphysik, Universit\"at Freiburg,
Hermann-Herder-Straße 3, 79104 Freiburg i.Br., Germany}

\date{\today} 
\begin{abstract}
We model coherent exciton transport in dendrimers by continuous-time
quantum walks (CTQWs). For dendrimers up to the second generation the
coherent transport shows perfect recurrences, when the initial excitation
starts at the central node. For larger dendrimers, the recurrence ceases
to be perfect, a fact which resembles results for discrete quantum
carpets. Moreover, depending on the initial excitation site we find that
the coherent transport to certain nodes of the dendrimer has a very low
probability. When the initial excitation starts from the central node, the
problem can be mapped onto a line which simplifies the computational
effort.  Furthermore, the long time average of the quantum mechanical
transition probabilities between pairs of nodes show characteristic
patterns and allow to classify the nodes into clusters with identical
limiting probabilities.  For the (space) average of the quantum mechanical
probability to be still or again at the initial site, we obtain, based on
the Cauchy-Schwarz inequality, a simple lower bound which depends only on
the eigenvalue spectrum of the Hamiltonian. 
\end{abstract}
\pacs{
71.35.-y, 
36.20.-r, 
36.20.Kd 
}
\maketitle

\section{Introduction}

In recent years dendrimers have been an active field of research, both
experimentally and theoretically, see, e.g.\, reference \cite{Voegtle} or
the Special Issue of Journal of Luminescence on the optical properties of
dendrimers.\cite{jlum05} Dendrimers are hyperbranched macromolecules with
a very regular structure.  Among a series of very interesting and
crossdisciplinary applications like drug delivery, dendrimers have been
theoretically investigated as light harvesting
antennae.\cite{mukamel1997,jiang1997,barhaim1998,flomenbom2005} Apart from
these theoretical works, there has been also a huge experimental effort to
probe transport
processes.\cite{kopelman1997,shortreed1997,lupton2001,varnavski2002,varnavski2002b}
Here, dendrimers are synthesized in a self-similar fashion by
hierarchically growing the dendrimer from a core.\cite{bharathi1995}
Depending on the site of the light absorbing states, the transport of
these excitations can be very efficient in some cases but also very
inefficient in others.  At high temperatures the transport is incoherent
and can be described by a hopping process.\cite{shortreed1997} However,
there is also experimental evidence for coherent interchromophore
transport processes within the
dendrimer.\cite{varnavski2002,varnavski2002b} When modeling the transport,
one usually assumes that the excitations are localized on the building
blocks of the dendrimer, i.e., either on the chromophores or the segments
connecting the chromophores. The latter leads to a different topology
usually referred to as being a (finite) Husimi cactus.\cite{poliakov1999}

There is a long-standing study of exciton transport in molecular
aggregates, not only in polymer physics \cite{Kenkre} but also in atomic
\cite{Shore} and in solid state physics.\cite{Davydov,Haken}  In solid
state physics, a dendrimer of infinite generation is known as the Bethe
lattice.  The incoherent exciton transport in dendrimers can be
efficiently modelled by random walks, see, for instance,
\cite{heijs2004,vlaming2005,blumen2005}. Here, the underlying topology of
the dendrimer determines the dynamics of the exciton motion and the
transport is described by a master (rate) equation. 

In this paper we will consider only the coherent transport, and we will
model the dynamics by Schr\"odinger's equation. Interestingly, this is
formally closely related to the master equation approach, where the
transfer over the system is given by the connectivity matrix ${\bf A}$ of
the dendrimer,\cite{mb2005a,mb2005b,mvb2005a,mb2006a} {\sl vide infra}.
This is also in close relation to H\"uckel's (or LCAO, linear combination
of atomic orbitals) theory,\cite{McQuarrie} where the elements of the
secular matrix are given by ${\bf A}$.

\section{Coherent exciton transport on graphs}

We model the coherent transport of excitons on graphs by so-called
continuous-time quantum walks (CTQWs) which are the quantum mechanical
analog of continuous-time random walks (CTRWs). 

A graph is a collection of connected nodes and to every graph there exists
a corresponding connectivity matrix ${\bf A} = (A_{ij})$, which is a
discrete version of the Laplace operator. The non-diagonal elements
$A_{ij}$ equal $-1$ if nodes $i$ and $j$ are connected by a bond and $0$
otherwise. The diagonal elements $A_{ii}$ equal the number of bonds which
exit from node $i$, i.e., $A_{ii}$ equals the functionality $f_i$ of the
node $i$.  

Classically, a CTRW is governed by a master equation for the conditional
probability, $p_{k,j}(t)$, to find the CTRW at time $t$ at node $k$ when
starting at node $j$.\cite{weiss,vankampen} The transfer matrix of the
walk, ${\bf T} = (T_{kj})$, is related to the connectivity matrix by ${\bf T}
= - \gamma {\bf A}$, where we assume for simplicity the transmission rate
$\gamma$ of all bonds to be equal and we set $\gamma\equiv 1$. 

CTQWs are obtained by identifying the Hamiltonian of the system with the
(classical) transfer operator (matrix), ${\bf H} = - {\bf T}$.\cite{farhi1998,childs2002,mb2005a} The whole accessible Hilbert space is
spanned by the basis vectors $|j\rangle$ associated with the nodes $j$ of
the graph.  A state $| j \rangle$ starting at time $t_0$ evolves in time
as $| j(t) \rangle = {\bf U}(t,t_0) |j \rangle$, where ${\bf U}(t,t_0) =
\exp(-i {\bf H} (t-t_0))$ is the quantum mechanical time evolution
operator (we have set $m\equiv1$ and $\hbar\equiv1$). 

The transition amplitude $\alpha_{k,j}(t)$ from state $| j \rangle$ at
time $0$ to state $|k\rangle$ at time $t$ reads then $\alpha_{k,j}(t) =
\langle k | {\bf U}(t,0) | j \rangle$ and obeys Schr\"odinger's equation.
Denoting the orthonormalized eigenstates of the Hamiltonian ${\bf H} =
-{\bf T}$ by $| q_n\rangle$, such that $\sum_n | q_n\rangle \langle  q_n |
= \boldsymbol 1$, the quantum mechanical transition probability is
\be
\pi_{k,j}(t) \equiv |\alpha_{k,j}(t)|^2 = \left| \sum_n \langle k| e^{-i
\lambda_n t} | q_n\rangle
\langle  q_n | j
\rangle \right|^2.
\label{qm_prob_full}
\ee
Note that classically $\sum_k p_{k,j}(t) = 1$, whereas quantum
mechanically $\sum_k |\alpha_{k,j}(t)|^2 =1$ holds.

\section{Topology \& connectivity of dendrimers}

\begin{figure}[htb]
\centerline{\includegraphics[clip=,width=\columnwidth]{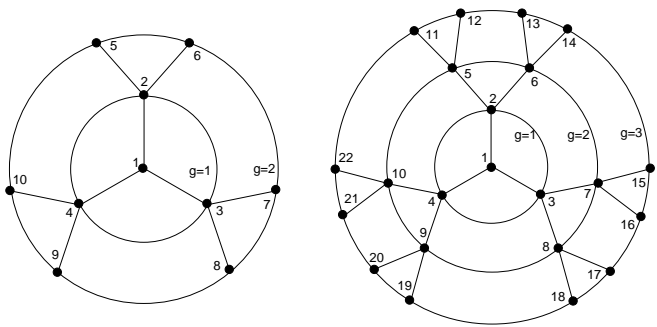}}
\caption{
Dendrimers of generation $G=2$ (left) and $G=3$ (right).
}
\label{topology}
\end{figure}

In the following we will consider dendrimers of generations $G$ whose
functionality is $3$, i.e.\ all internal nodes of the dendrimer have $3$
bonds, whereas the outermost nodes have only one bond.  The stucture of
such dendrimers is exemplified in Fig.~\ref{topology} for dendrimers of
generations $G=2$ and $G=3$.  Note that the number of nodes belonging to
the $g$-th generation ($g\geq1$) is $3\cdot2^{g-1}$ and that it grows
exponentially with $g$. Moreover, the total number of nodes in the
dendrimer of generation $G$ is $N=3\cdot2^G-2$.

The connectivity matrix of these dendrimers has a very simple structure.
One has $A_{ii}=3$ for all the nodes in the first $G-1$ generations and
$A_{ii}=1$ for the nodes in generation $G$. The bonds are represented by
the off-diagonal matrix elements $A_{ij}$. Here, every node in generation
$g\ge1$ is connected to two consecutively numbered nodes in generation $g+1$
and to one node in generation $g-1$.

The eigenmodes of such dendrimers were studied in.\cite{cai1997} There,
for the dendrimers of generations $G=1$ and $G=2$, the eigenvalues and
eigenvectors of ${\bf A}$ were explicitly calculated.  The eigenvectors
determine the eigenmodes of the dendrimer, see e.g.\ Fig.~3 of
\cite{cai1997}. It was further shown that there are $G+1$ nondegenerate
eigenvalues, one of which is always $\lambda_0=0$. We used these results
to check that our numerical diagonalization for the small dendrimers is
correct.

\section{Transport on dendrimers}

In the following section we present results for coherent exciton transport
on dendrimers. The results were obtained by numerically determining the
eigenvalues and eigenvectors of the corresponding connectivity matrix,
using the standard software package MATLAB. 

\subsection{Dynamics of an excitation starting at the center}

\begin{figure}[htb]
\centerline{\includegraphics[clip=,width=\columnwidth]{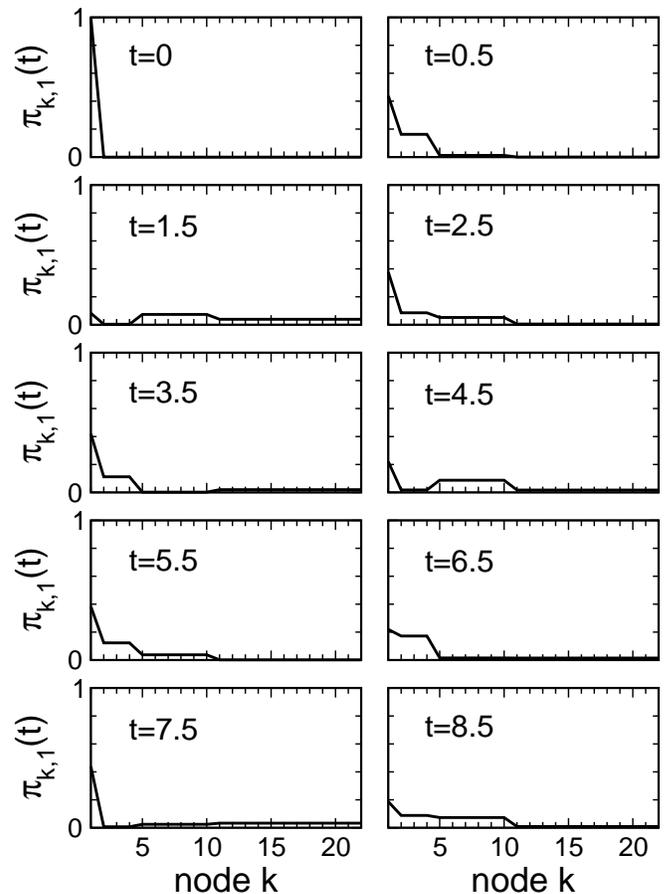}}
\caption{
Snapshots of the transition probability for a $G=3$ dendrimer at different times $t$.  The initial excitation is at the central node.
}
\label{d22_anregung_innen}
\end{figure}

We start by focusing on an excitation which starts at the central node
$1$.  For a $G=3$ dendrimer, Fig.~\ref{d22_anregung_innen} shows snapshots
of the temporal development of transition probabilities.  The nodes are
numbered according to Fig.\ref{topology}. In this way the nodes $2-4$
belong to generation $g=1$, the nodes $5-10$ to $g=2$, and the nodes
$11-22$ to $g=3$. Here, the excitation spreads over the whole
dendrimer and already after a short period of time, $t\approx1.5$, there
is a considerable probability to be at the outermost nodes, which is
larger than the probabilities to be at nodes belonging to other
generations. The initial ``speed'' of the exciton is calculated based on
the number of bonds (i.e.\ the chemical distance) between the initial node
and one of the outermost nodes, divided by the time to reach that node.
Here, we have $3/1.5=2$. As we will show below, the dynamics of an
excitation starting at the central node $1$ over the dendrimer can be
mapped onto a line. For an infinite line, the transition probabilities can
be expressed by Bessel functions of the first kind.\cite{mb2005b} From
this the speed can be calculated; it is at all times equal to $2$. For
finite lines, the initial speed is still equal to $2$. Thus our results
here are in accordance with what can be expected for waves propagating
through a regular graph.

Remarkably, for the $G=2$ dendrimer the transition probabilities are fully
periodic when the coherent excitation starts from the central node $1$
(the same holds for the $G=1$ dendrimer, too).  In this case we obtain,
based on the analytically determined eigenvalues and eigenvectors that all
$\pi_{k,1}(t)$ have the following, periodic form:
\bea
&&\pi_{k,1}(t) = \left| \sum_n e^{-i\lambda_nt} \langle k | q_n \rangle
\langle q_n | 1 \rangle \right |^2 \nonumber \\
&& = a^{(k)}_0+ a^{(k)}_1 \cos(2t) 
+ a^{(k)}_2 \cos(3t) + a^{(k)}_3 \cos(5t) .
\label{pik1_d10}
\eea
Note that due to rotational symmetry, the transition probabilities from
the central node to nodes belonging to the same generation are equal.
Because of this we only have to list three different transition
probabilities.  We hence choose, exemplarily, the nodes $1$, $2$, and
$10$, for which we find the coefficients
\bea
&& a^{(1)}_0 = \frac{21}{50} \ , \ a^{(1)}_1
= \frac{1}{10} \ , \ a^{(1)}_2 = \frac{2}{5}  \ , \ a^{(1)}_3 = \frac{2}{25}
\nonumber \\
&& a^{(2)}_0 = \frac{49}{450}\ , \ a^{(2)}_1
= \frac{1}{30} \ , \ a^{(2)}_2 = -\frac{4}{45}  \ , \ a^{(2)}_3 = -\frac{4}{75} 
\nonumber \\
&& a^{(10)}_0 = \frac{19}{450}\ , \
a^{(10)}_1
= -\frac{1}{30} \ , \ a^{(10)}_2 = -\frac{1}{45}  \ , \ a^{(10)}_3 =
\frac{1}{75}. 
\nonumber
\eea

If follows that there is a perfect revival of the initial state for each
$t=2n\pi$ where $n\in\mathbb{N}$. At $t=(2n-1)\pi$ the main part of the
excitation is equally distributed among the nodes of the outermost
generation. This revival of the initial probability distribution resembles
results obained for continuous \cite{kinzel1995,fgrossmann1997} and
discrete quantum carpets.\cite{iwanow2005,mb2005b} For discrete quantum
carpets, the revival is only perfect for small cycles of length
$N=1,2,3,4,$ and $6$. When the cycles become larger, there are only
partial revivals of the initial state. The same can be observed for
dendrimers of generation $G\geq3$, as will become clear in the following.

\subsection{Mapping onto a line}

Due to the rotational symmetry, the dynamical problem of a dendrimer
initially excited at its center $1$ can be mapped onto a linear segment.
In the classical, CTRW case, this result has been widely used, see
Ref.~\cite{barhaim1998}.  For a CTQW on a graph consisting of two Cayley
trees this has been done in, e.g., \cite{childs2002,mb2005a}.  For each
generation with $g\ge1$ we introduce linear combinations of states
\be
|g\rangle \equiv \frac{1}{\sqrt{3\cdot 2^{g-1}}} \sum_{k\in g} |k\rangle.
\ee
The state of generation $0$, $|g_0\rangle$, is identical to that of the
central node $|1\rangle$. A CTQW through the generations is now governed
by a new Hamiltonian $\tilde{\bf H}$ which is defined by
\be
\tilde H_{gg'} = \langle g | {\bf H} | g' \rangle.
\ee
Given ${\bf H}$ and the construction of the generation states $|
g\rangle$, $\tilde{\bf H}$ is a real and tridiagonal matrix whose elements
obey $\tilde H_{gg'} = \tilde H_{g'g}$ and are given by
\bea
\tilde H_{gg} &=& \langle g | {\bf H} | g \rangle = f_g \ ,\\
\tilde H_{g,g\pm1} &=& \langle g | {\bf H} | g\pm1 \rangle 
= \frac{\sqrt{2^{\mp1}}}{3\cdot2^{g-1}} \sum_{k \in g}\sum_{j\in
g\pm1} \langle k | {\bf H} | j \rangle
\nonumber \\
&=& \begin{cases} -\sqrt{3} & \mbox{for} \quad \tilde H_{01} = \tilde H_{10}\\
-\sqrt{2} & \mbox{else}. \end{cases}
\eea
Here, $f_g$ is the functionality of the nodes in generation $g$ and all
other elements of $\tilde{\bf H}$ are zero. In this way the original
eigenvalue problem which grows exponentially with $G$ has been reduced to
an eigenvalue problem which grows only linearly with $G$.  The $G+1$
eigenvalues of the new matrix $\tilde{\bf H}$ are the nondegenerate
eigenvalues of the original matrix ${\bf H}$. In terms of eigenmodes,
these are the modes of the full dendrimer, where, in a classical
picture,\cite{cai1997} nodes belonging to the same generation move in the
same direction.  

Now, the transition amplitude between different generations can be written
as $\langle g | \exp(-i\tilde{\bf H}t) | g_0 \rangle$ .  It is easy to
show that the transition probabilities for a line of $3$ nodes obtained
from the $G=2$ dendrimer are, see Eq.(\ref{pik1_d10}),
\be
\left|\langle g| \exp(-i{\bf H}t) | g_0\rangle\right|^2 =
\sum_{k\in g} \pi_{k,1}(t).
\ee

\begin{figure}[htb]
\centerline{\includegraphics[clip=,width=\columnwidth]{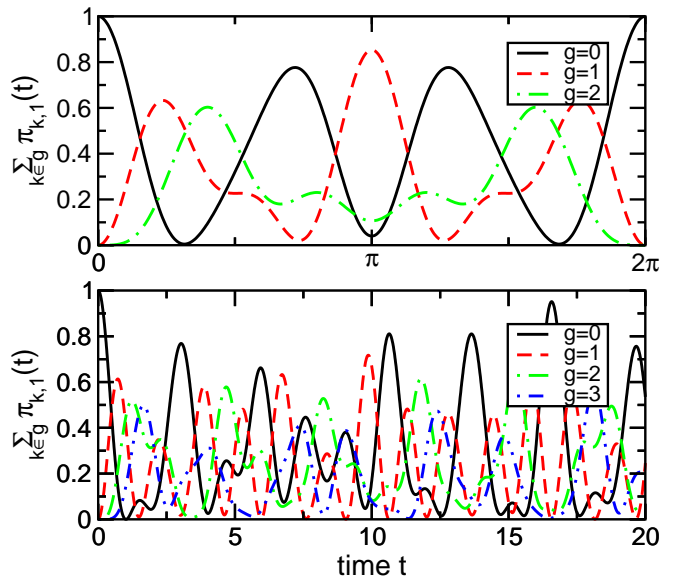}}
\caption{
(Color online). Cumulative transition probabilities that a coherent excitation starting at the central node is at time $t$ at a node of generation $g$.  We present the situation for dendrimers of generations (a) $G=2$ and (b) $G=3$. Note the complete revival at $t=2\pi$ for the $G=2$ dendrimer.  
}
\label{d22_anregung_innen_gen}
\end{figure}

The cumulative transition probabilities for the $G=2$ and the $G=3$
dendrimer to go from one generation of the dendrimer to another one when
starting at $g=0$ are shown in Fig.~\ref{d22_anregung_innen_gen}. For
generations $G\ge3$, we do not find perfect revivals of the initial
probability anymore.

\subsection{Dynamics of excitations starting off-center}

\begin{figure}[htb]
\centerline{\includegraphics[clip=,width=\columnwidth]{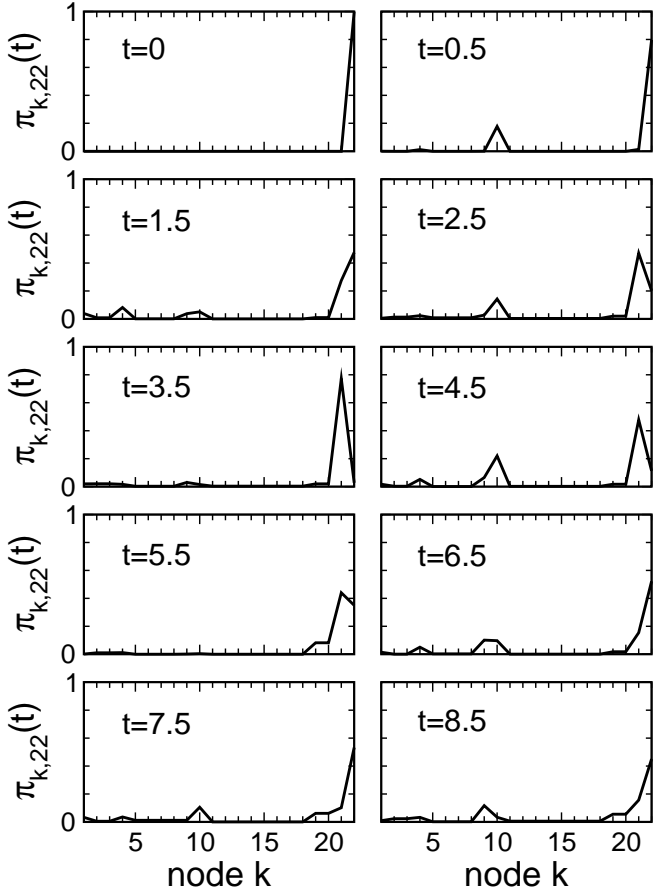}}
\caption{
Snapshots of the transition probability for a $G=3$ dendrimer at different times $t$.  The excitation starts at the outermost node $22$.
}
\label{d22_anregung_aussen}
\end{figure}

If the initial excitation is placed at one of the nodes of the outermost
generation $g=G$ of the dendrimer, the picture changes. Snapshots of the
transition probability for a $G=3$ graph where the excitation starts at
the outermost node $22$ are shown in Fig.~\ref{d22_anregung_aussen}.  Even
classically the propagation through the dendrimer gets to be much slower
than in the previous case, see e.g.\
\cite{barhaim1997,rana2003,heijs2004}. Nevertheless, eventually the
excitation will classically propagate through the whole graph and in the
long time limit the probability will be equipartitioned among all nodes. 

Quantum mechanically this effect is even more dramatic. For the times
shown in Fig.~\ref{d22_anregung_aussen}, the main fraction of
$\pi_{k,j}(t)$ stays in a small region closely connected by bonds to the
initial node $j$, and the transfer to other sites is highly unlikely.  For
the $G=3$ dendrimer shown in Fig.~\ref{topology}, the excitation is
located with very high probability on the initial node $22$ and on the
nodes $21$ and $10$. On the time scales shown in
Fig.~\ref{d22_anregung_aussen}, the probability for nodes outside the
branch (consisting of the nodes $1,4,9,10,19,20,21,$ and $22$) to be
excited is very low . 

As we show in the following section, also at long times the limiting
probability for the excitation to reach the other branches of the
dendrimer stays very low. 

\subsection{Long time limit}\label{long_time}

Quantum mechanically the temporal development is symmetric to inversion;
this prevents $\pi_{k,j}(t)$ from having a definite limit for
$t\to\infty$. In order to compare the classical long time probability with
the quantum mechanical one, one usually uses the limiting probability (LP)
distribution \cite{aharonov2001}
\be
\chi_{k,j} \equiv \lim_{T\to\infty} \frac{1}{T} \int_0^T dt \
\pi_{k,j}(t),
\label{limprob}
\ee
which can be rewritten by using the orthonormalized eigenstates of the
Hamiltonian, $| q_n\rangle$, as \cite{mvb2005a}
\bea
\chi_{k,j} &=& 
\lim_{T\to\infty}\frac{1}{T} \int_0^T dt \ \left|
\sum_{n}\langle k | e^{-i{\bf H}t} |
q_n \rangle \langle q_n |
j \rangle \right|^2 
\nonumber \\
&=&
\sum_{n,m} 
\delta_{\lambda_n,\lambda_m}
\langle k |
q_n \rangle \langle q_n |
j \rangle \langle j | q_m
\rangle \langle q_m | k \rangle. 
\label{limprob_ev}
\eea
Some eigenvalues of $\bf H$ might be degenerate, so that the sum in
Eq.~(\ref{limprob_ev}) can contain terms belonging to different
eigenstates $| q_n\rangle$ and $| q_m\rangle$.  

We can use the Cauchy-Schwarz inequality \cite{abramowitz}
\be
\left| \int_0^T dt \
f(t) g(t) \right|^2 \le \left(\int_0^T dt \ |f(t)|^2\right) \left(\int_0^T
dt \ |g(t)|^2
\right)
\ee
to obtain a lower bound for the LP. With $f(t)\equiv \alpha_{k,j}(t)$ and
$g(t)\equiv1$, the time integral in Eq.~(\ref{limprob}) fulfills the
inequality
\be
\int_0^T dt \ |\alpha_{k,j}(t)|^2
\ge \frac{1}{T}\left| \int_0^T dt \
\alpha_{k,j}(t) \right|^2 .
\label{csi_alpha}
\ee
This results in
\be
\chi_{k,j} \ge \lim_{T\to\infty} \left| \frac{1}{T} \int_0^T dt \
\sum_n \langle k| e^{-i
\lambda_n t} | q_n\rangle
\langle  q_n | j
\rangle
\right|^2.
\label{csi_alpha1}
\ee
The only term in the sum over $n$ in Eq.~(\ref{csi_alpha1}) which survives
after integration and taking the limit $T\to\infty$ is the one with
$\lambda_0=0$.  In terms of eigenmodes, the eigenstate corresponding to
$\lambda_0$ is the one for which in a classical picture the dendrimer
moves as a whole in one direction. The corresponding eigenvector can be
written as $|q_0\rangle = 1/\sqrt{N}\sum_{j=1}^{N} | j\rangle$. Since the
states $|j\rangle$ form a complete, ortho-normalized basis set, i.e.\
$\langle k | j \rangle = \delta_{kj}$, we get $\langle k | q_0 \rangle =
\langle q_0 | j \rangle = 1/\sqrt{N}$ and therefore, with
Eq.~(\ref{csi_alpha1}),
\be
\chi_{k,j} \ge \left| \langle k | q_0 \rangle \langle q_0 | j \rangle
\right|^2 = \frac{1}{N^2}.
\label{lbchi}
\ee

Classically, the LP is equipartitioned among all the nodes, i.e.\
$\lim_{t\to\infty} p_{k,j}(t) = 1/N$ for all nodes. Quantum mechanically
this is not the case. However, for regular graphs like rings and for
initial conditions localized on one node, the quantum mechanical LPs are
{\sl almost} equipartitioned, i.e., there are only a few nodes whose LPs
differ from the rest.\cite{mb2006a} For more complex structures, the LPs
display large variations, depending on the nodes, as we proceed to
demonstrate.

\begin{figure}[htb]
\centerline{\includegraphics[clip=,width=\columnwidth]{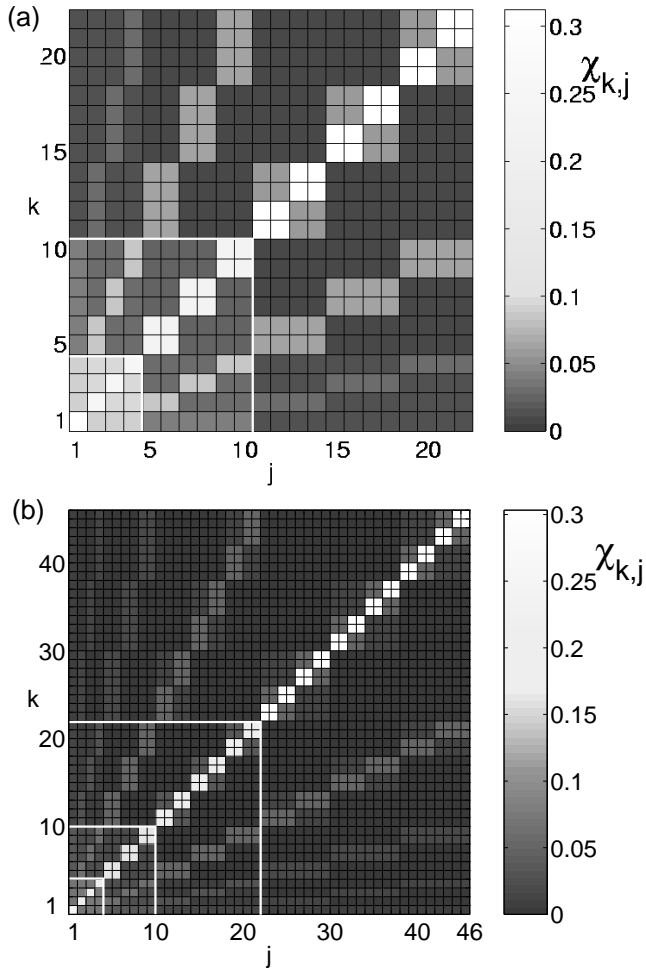}}
\caption{
Limiting probabilities for dendrimers with (a) $G=3$ and (b) $G=4$. The white lines indicate the limiting distributions for the dendrimers of generations smaller than $G$.  
}
\label{limiting}
\end{figure}

Figure \ref{limiting} shows the LPs $\chi_{k,j}$ as a contour plot where
the axes are labeled by the nodes $k=1,\cdots,N$ and $j=1,\cdots,N$.
Bright shadings correspond to high values of the LP, whereas dark shadings
correspond to low LPs.  The diagonal has high values, meaning that
excitations starting at any node $j$ have a high LP to be at node $j$
again.  Note that all values of $\chi_{k,j}$ are clearly larger than the
lower bound given in Eq.~(\ref{lbchi}), which is $(1/22)^2$ for the $G=3$
dendrimer and $(1/46)^2$ for the $G=4$ dendrimer.

The structures of the LP distributions of dendrimers are self-similar
generation after generation.  The white lines in Fig.~\ref{limiting} are a
guide to the eye, indicating this fact.

Furthermore, different nodes $k$ and $l$ may have the same LP,
$\chi_{k,j}=\chi_{l,j}$. We hence combine all the nodes having (up to our
numerical precision, $10^{-10}$) the same LP into a cluster. Note, however,
that the separation of the nodes into clusters depends on the initially
excited node, namely on $j$. For an excitation starting at the center
(node $1$), the clusters correspond exactly to the different generations
of the dendrimer. In the general case, when starting from a non-central
node, we still find from Fig.~\ref{limiting} that nodes
belonging to the same cluster also belong to the same generation (the
converse is not necessarily true).  

\begin{figure}[htb]
\centerline{\includegraphics[clip=,width=\columnwidth]{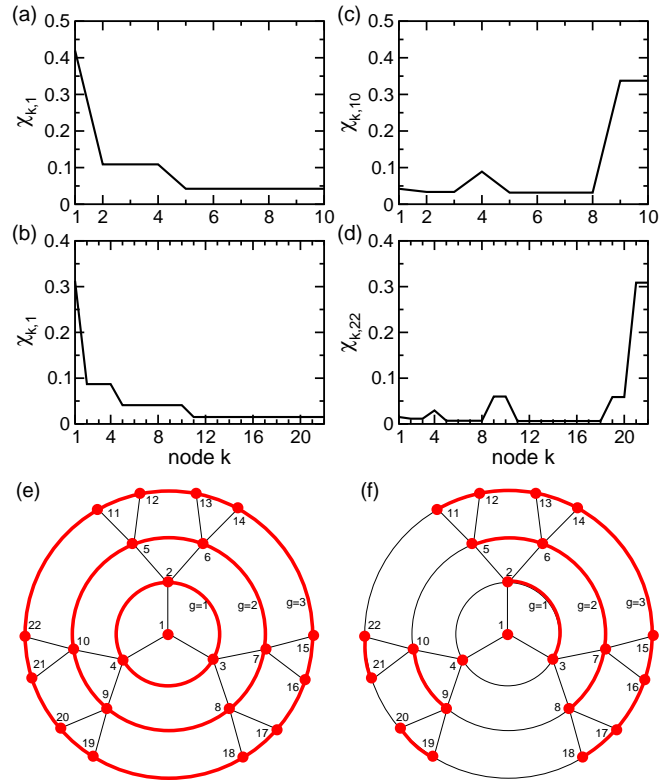}}
\caption{
(Color online). Upper two panels: Limiting probabilities for the $G=2$ (top row) and the $G=3$ (bottom row) dendrimer when the excitation starts from the central node $1$ [(a) and (b)] and from one of the outermost nodes [(c) from node $10$ and (d) from node $22$].\\ Lower panel: Groups of nodes (clusters) having the same limiting probability $\chi_{k,j}$ for a dendrimer of generation $G=3$. In (e) the excitation starts from node $1$ and in (f) from node $22$.  Nodes connected by thick (red) lines belong to the same cluster.
}
\label{limiting_cut}
\end{figure}

We now focus on the two special cases of initial excitations given in the
previous sections. Figure~\ref{limiting_cut} presents the LP distributions
for the $G=2$ and $G=3$ dendrimers; in both cases the excitation starts
either at the central node $1$ or at one of the outermost nodes (node $10$
for $G=2$ and node $22$ for $G=3$). 

Having the initial excitation at the central node results in a totally
symmetric LP distribution, i.e.\ in each generation all nodes have the
same LP. In this case we can again simplify the problem by considering
only the cumulative probabilities. The LP given in Eq.~(\ref{limprob_ev})
simplifies, because the cumulative probabilities contain only
nondegenerate eigenvalues, i.e.\
\be
\chi_{g,g'} = \sum_n \left|\langle g|\tilde q_n\rangle\right|^2
\left|\langle \tilde q_n | g'\rangle\right|^2.
\ee 
Here, the $|\tilde q_n\rangle$ are the eigenstates of the reduced
Hamiltonian $\tilde{\bf H}$, which does not have degenerate eigenvalues.
For the totally recurrent case of an excitation starting at the center of
the $G=2$ dendrimer, the LP distribution follows directly from
Eq.~(\ref{pik1_d10}) as
\be
\chi_{k,1} = a_0^{(k)}.
\ee
Note that for an excitation starting at the center the number of different
clusters equals $G+1$.

For an excitation starting from one of the outermost nodes, the LPs are
less regular. However, also in this case the LPs cluster.
Figure~\ref{limiting_cut}(f) shows the situation for the $G=3$ dendrimer
when starting from node $j=22$. Here, nodes belonging to the same cluster
are connected by thick (red) lines. The cluster structure beyond the
central node $1$ can be understood on simple symmetry grounds. The
structure on the side of the initial node, however, is more complex. Here,
nodes $22$ and $21$ have the same LP and form one cluster. The same holds
(remarkably) for nodes $9$ and $10$ (another cluster) and for nodes $19$
and $20$ (yet another cluster).

\begin{figure}[htb]
\centerline{\includegraphics[clip=,width=0.5\columnwidth]{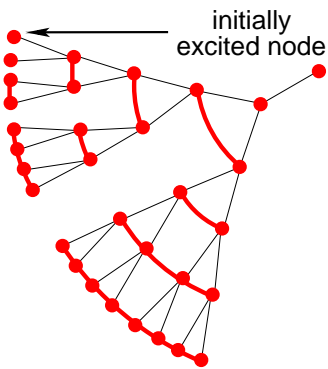}}
\caption{
(Color online). Clusters of the same limiting probability $\chi_{k,j}$ in the branch with the initial excitation for a dendrimer of generation $G$. Nodes connected by thick (red) lines belong to the same cluster.
}
\label{limiting_clust}
\end{figure}

For larger dendrimers, while the general cluster pattern is preserved,
some details change.  Figure~\ref{limiting_clust} shows the situation for
a dendrimer of dimension $G=5$, for an excitation starting at a peripheral
node.  Again we indicate clusters by connecting nodes by thick (red)
lines.  A change to be noticed is that for $G\ge5$ the initially excited
node does not form anymore a cluster with its next-nearest node of the
same generation. It appears as if such two nodes belong to the same
cluster only when the dendrimer has $G\le4$. Thus the total number of
clusters is $N_C \equiv (G^2+G+6)/2$ for $G\geq5$ and $N_C-1$ for
$G\leq4$. Moreover, the study of the dendrimer with $G=6$ shows an
analogous situation to our findings for $G=5$.

\subsection{Averaged probabilities}

One interesting question to ask is, what is the probability,
$\pi_{j,j}(t)$, to be at the initial site after some time $t$? As shown
above, for the $G=2$ dendrimer $\pi_{k,1}(t)$ and hence $\pi_{1,1}(t)$ are
periodic.  However, for larger dendrimers and/or different initial
conditions this does not have to be the case. 

Classically, there exists a simple expression for the {\sl average}
probability to be still or again at the initially excited node. One has
(see e.g.\ Ref.~\cite{blumen2005}):
\be
\overline{p}(t) \equiv \frac{1}{N} \sum_{j=1}^{N} p_{j,j}(t) =
\frac{1}{N} \sum_{n=1}^{N} \ \exp\big(-\lambda_n t\big).
\ee
This result is quite remarkable, since it depends only on the eigenvalue
spectrum of the connectivity matrix ${\bf A}$ but {\sl not} on the
eigenvectors.

\begin{figure}[htb]
\centerline{\includegraphics[clip=,width=\columnwidth]{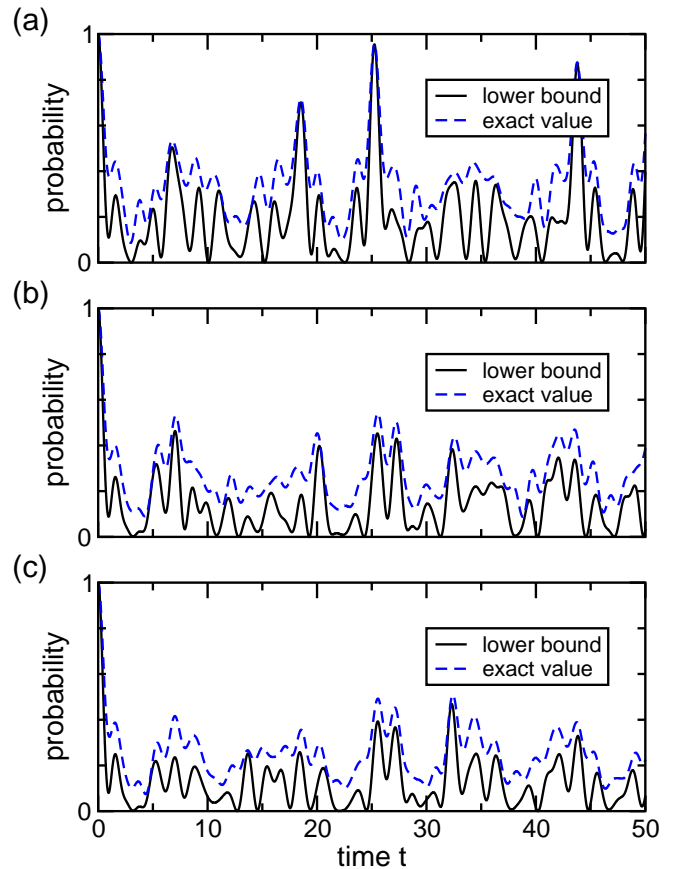}}
\caption{
(Color online). Comparison of the averaged probability to be still or again at the initial node with the lower bound given in Eq.~(\ref{csi}) for generations (a) $G=2$, (b) $G=3$, and (c) $G=4$.
}
\label{csu_g10_bis_d46}
\end{figure}

Quantum mechanically, the average is given by
\be
\overline{\pi}(t) \equiv \frac{1}{N} \sum_{j=1}^{N} \pi_{j,j}(t).
\ee
By using the Cauchy-Schwarz inequality we obtain a lower bound for
$\overline{\pi}(t)$,
\be
\overline{\pi}(t) 
= 
\frac{1}{N} \sum_{j=1}^{N} |\alpha_{j,j}|^2 \frac{1}{N} \sum_{l=1}^{N} 1
\geq \left| \frac{1}{N} \sum_j \alpha_{j,j} \right|^2 .
\ee
With Eq.~(\ref{qm_prob_full}) we have
\be
 \sum_j \alpha_{j,j} = \sum_j \sum_n  e^{-it
\lambda_n} \langle q_n | j \rangle
\langle  j| q_n
\rangle = \sum_n  e^{-it
\lambda_n}, 
\ee
and therefore 
\be
\overline{\pi}(t) \geq 
\frac{1}{N^2} \sum_{n,m} \ \exp\big[-i(\lambda_n - \lambda_m)t\big].
\label{csi}
\ee
In analogy to the classical case, the lower bound in Eq.~(\ref{csi})
depends only on the eigenvalues and {\sl not} on the eigenvectors of ${\bf
A}$. Note that for a CTQW on a simple regular network with periodic
boundary conditions the lower bound in Eq.~(\ref{csi}) becomes
exact.\cite{bbm2006a,vmb2006a} 

Figure \ref{csu_g10_bis_d46} shows the exact value of the average
probability $\overline{\pi}(t)$ as well as the lower bound given in
Eq.~(\ref{csi}) for dendrimers of three different generations. One notes
immediately that the fluctuations of the lower bound are larger that the
ones of the exact value. However, the qualitative agreement between the
two curves is remarkable: The positions in time of the extrema of the
lower bound and of the exact curve almost coincide.  Especially the strong
maxima of $\overline{\pi}(t)$ are well reproduced by the lower bound.

In fact, fluorescence spectroscopy experiments like, e.g., the ones of
Varnavski {\sl et al.}\cite{varnavski2002,varnavski2002b} allow, in
principle, to distinguish whether the transport is rather classical or
rather quantum mechanical.  In order to identify excitonic coherence,
Lupton {\sl et al.} compared a coupled harmonic oscillator model to
photoluminescence spectra.\cite{lupton2002b} Another suggestion is the one
by Heijs {\sl et al.}, in which pump-probe and fluorescence experiments
determine the efficiency of the excitation transfer,\cite{heijs2004} from
which one can infer the underlying transport mechanism. We close by noting
that such models are, as a rule, quite idealized, so that one must be
careful in relating the models to the experimental findings.

\section{Conclusions}

In conclusion, we have modelled the coherent exciton transport by
continuous-time quantum walks. The transport is exclusively determined by the
topology of the dendrimer, i.e.\ by its connectivity matrix. For the $G=2$
dendrimer the transport is completely recurrent when the central node is
initially excited. In this case the quantum mechanical transition
probabilities are fully periodic. For larger dendrimers and/or different
initial conditions we observe only partial recurrences.

To compare these results to those of the classical (incoherent) case, we
calculated the long time average of the transition probabilities.
Depending on the initial conditions, these show characteristic patterns.
Furthermore, the limiting probability distributions can be characterized
by clusters of nodes having the same limiting probabilities.

Furthermore, we calculated a lower bound for the average probability to be
still or again at the initial node after some time $t$. This lower bound
depends only on the eigenvalues of ${\bf A}$ and agrees qualitatively well with
the exact value; especially the maxima of $\overline{\pi}(t)$ are well
reproduced by the lower bound. 

\section*{Acknowledgments}

This work was supported by a grant from the Ministry of Science, Research
and the Arts of Baden-W\"urttemberg (Grant No.\ AZ: 24-7532.23-11-11/1).
Further support from the Deutsche Forschungsgemeinschaft (DFG) and the
Fonds der Chemischen Industrie is gratefully acknowledged.

\end{document}